\documentclass[nonacm,sigplan,screen]{acmart}

\usepackage{filecontents}
\usepackage{algorithm}
\usepackage{algpseudocode}
\usepackage{multirow}
\usepackage{makecell}
\usepackage{fancyhdr}
\usepackage{listings}
\usepackage{balance}

\lstdefinestyle{mystyle}{
  basicstyle=\ttfamily\footnotesize,   % 代码字体
  keywordstyle=\color{blue},           % 关键字颜色
  commentstyle=\color{gray},           % 注释颜色
  stringstyle=\color{red},             % 字符串颜色
  numberstyle=\tiny\color{gray},       % 行号样式
  breaklines=true,                     % 代码自动换行
  breakatwhitespace=false,             % 不仅限于空格换行
  frame=single,                        % 添加边框
  captionpos=b,                        % 标题位置（b=bottom, t=top）
  numbers=left,                        % 行号位置
  numbersep=8pt,                       % 行号与代码间隔
  tabsize=2,
  columns=flexible,
  showstringspaces=false               % 不显示字符串中的空格
}

\setcopyright{none}
\acmBooktitle{}
\ccsdesc[500]{Computing methodologies~Machine learning}
\ccsdesc[500]{Computing methodologies~Parallel computing methodologies}

\keywords{LLM Inference, Scheduling Optimization}

\begin{document}

\title{Past-Future Scheduler for LLM Serving under SLA Guarantees}

\author{Ruihao Gong}
\authornote{Equal contribution.}
\affiliation{
  \institution{Beihang University}
  \city{Beijing}
  \country{China}
}
\email{gongruihao@buaa.edu.cn}

\author{Shihao Bai}
\authornotemark[1]
\affiliation{
  \institution{SenseTime}
  \city{Shanghai}
  \country{China}
}
\email{baishihao@sensetime.com}

\author{Siyu Wu}
% \orcid{0009-0006-4295-9983}
\authornotemark[1]
\affiliation{
  \institution{Beihang University}
  \city{Beijing}
  \country{China}
}
\email{wusiyu@buaa.edu.cn}

\author{Yunqian Fan}
\affiliation{
  \institution{SenseTime}
  \city{Shanghai}
  \country{China}
}
\email{fanyunqian@sensetime.com}

\author{Zaijun Wang}
\affiliation{
  \institution{SenseTime}
  \city{Shanghai}
  \country{China}
}
\email{wangzaijun@sensetime.com}

\author{Xiuhong Li}
\affiliation{
  \institution{Peking University}
  \city{Beijing}
  \country{China}
}
\email{lixiuhong@pku.edu.cn}

\author{Hailong Yang}
\authornote{Corresponding author.}
\affiliation{
  \institution{Beihang University}
  \city{Beijing}
  \country{China}
}
\email{hailong.yang@buaa.edu.cn}

\author{Xianglong Liu}
\authornotemark[2]
\affiliation{
  \institution{Beihang University}
  \city{Beijing}
  \country{China}
}
\email{xlliu@buaa.edu.cn}

% for header
% \renewcommand{\shortauthors}{R. Gong, S. Bai, S. Wu, et al.}
\renewcommand{\shortauthors}{Ruihao Gong et al.}
%% No italics, no superscripts
%% Use footnote or author note to identify equal contribution and/or contact author info

\begin{abstract}
The exploration and application of Large Language Models (LLMs) is thriving. To reduce deployment costs, continuous batching has become an essential feature in current service frameworks. The effectiveness of continuous batching relies on an accurate estimate of the memory requirements of requests. However, due to the diversity in request output lengths, existing frameworks tend to adopt aggressive or conservative schedulers, which often result in significant overestimation or underestimation of memory consumption. Consequently, they suffer from harmful request evictions or prolonged queuing times, failing to achieve satisfactory throughput under strict Service Level Agreement (SLA) guarantees (a.k.a. goodput), across various LLM application scenarios with differing input-output length distributions.
To address this issue, we propose a novel Past-Future scheduler that precisely estimates the peak memory resources required by the running batch via considering the historical distribution of request output lengths and calculating memory occupancy at each future time point. It adapts to applications with all types of input-output length distributions, balancing the trade-off between request queuing and harmful evictions, thereby consistently achieving better goodput. Furthermore, to validate the effectiveness of the proposed scheduler, we developed a high-performance LLM serving framework, LightLLM, that implements the Past-Future scheduler. Compared to existing aggressive or conservative schedulers, LightLLM demonstrates superior goodput, achieving up to 2-3$\times$ higher goodput than other schedulers under heavy loads. LightLLM is open source to boost the research in such direction (\href{https://github.com/ModelTC/lightllm}{https://github.com/ModelTC/lightllm}).

\end{abstract}

\maketitle % should come after the abstract
% \pagestyle{plain} % should come right after \maketitle

% Intro Figure
% 

\begin{figure}[t]
  \centering
  \includegraphics[width=\linewidth]{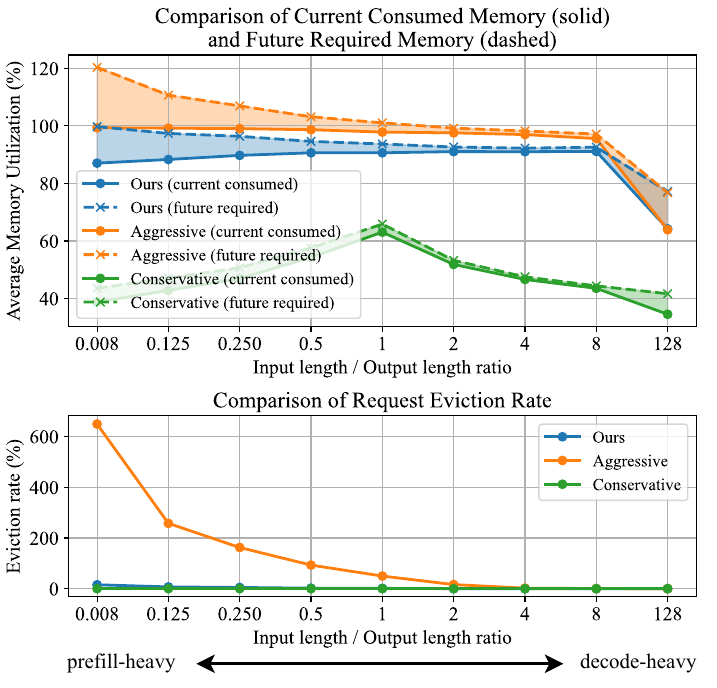}
  \caption{Comparison of current consumed memory (solid), future required memory (dashed), and request eviction rate (>100\% means one request is evicted multiple times on average) for different schedulers under various input-output length distributions. Lower memory utilization indicates lower efficiency, while future required memory exceeds the memory capacity will lead to request evictions afterward.}
  \label{fig:intro}
\end{figure}

\section{Introduction}

% Background

The burgeoning field of Large Language Models (LLMs) marks a watershed in technology, vastly enhancing potential applications across diverse domains such as chatbots\cite{bard,chatgpt}, summarization~\cite{see-etal-2017-get,paulus2018a,nallapati-etal-2016-abstractive}, code generation~\cite{copilot}, and advanced reasoning~\cite{wei2022chain,wei2022emergent}. Models such as GPT-4~\cite{openai2023gpt4}, Claude~\cite{claude}, and Llama~\cite{touvron2023llama} exemplify this trend, harnessing hundreds of billions of parameters for exceptional performance. However, the above breakthroughs come with high deployment expenses, especially under Service Level Agreement (SLA) guarantees. SLA guarantees, which constrain the first response time of requests and the maximum response interval between outputs, are crucial to user experience. 
For real-world LLM services, throughput under SLA guarantees, also known as goodput, is the well-adopted metric that accurately reflects the system's throughput capacity. Take ChatGPT for example, with 100 million weekly active users, it strives to maintain stable and scalable services, even with a robust infrastructure of 10,000 GPUs. In such a context, optimizing system goodput with limited hardware resources is not only advantageous but also essential, which helps to reduce cost and expand the reach of LLMs.

To reduce costs and enhance performance, various innovative approaches have been developed, each targeting different aspects of the system. The primary focus of these efforts is to improve the inference backend and optimize the serving schedule. For improving the inference backend, existing works such as FlexGEN~\cite{sheng2023flexgen} demonstrate the power of parallel processing. Using pipeline parallelism, FlexGEN significantly increases inference throughput on lower-end processing units. Other works such as Flash Attention~\cite{dao2022flashattention}, Flash Decoding, and their advanced iteration, Flash Attention V2~\cite{dao2023flashattention2} and Flash Decoding++~\cite{hong2023flashdecoding++}, have been devoted in creating high-performance kernels. These works are particularly focused on optimizing the Attention module, a core component of LLMs. In addition, GPTQ~\cite{gptq} and OS+~\cite{osplus} utilize quantization to accelerate the inference. Furthermore, PageAttention ~\cite{kwon2023efficient} is a notable development for memory management in LLM, designed to prevent memory fragmentation during request serving.
For optimizing the serving schedule, the scheduling strategies are crucial for system throughput. Traditional fixed batch size scheduling is inadequate for LLMs, which cannot handle varying input lengths and generate unpredictable output sizes, leading to unnecessary response delay. Continuous batching~\cite{yu2022orca} has become a key component of current frameworks to address the above limitation by adjusting the batching granularity at each request to each decoding step, thereby reducing the memory and computation waste. The key to the success of continuous batching lies in accurately estimating the memory requirements of the requests. To estimate the request memory requirement, the existing request schedulers can be classified into two categories: 1) the conservative scheduler that overestimates memory consumption, and 2) the aggressive scheduler that underestimates memory consumption.

Despite the performance achieved by existing schedulers, they still struggle to achieve satisfactory goodput across diverse LLM service scenarios. For example, both conservative and aggressive schedulers can easily breach SLA constraints under high concurrency and thus lead to low goodput with poor user experience. As shown in Figure \ref{fig:intro}, the conservative schedulers~\cite{tgi, fastgen} often queue requests based on the assumption that each request will generate up to a pre-set maximum number of new tokens which rarely occurs in practice. In addition, they overlook that the requests in the running batch will dynamically finish at different future time points, and thus lead to considerably low memory utilization and long first response latency. On the other hand, the aggressive schedulers~\cite{kwon2023efficient} attempt to incorporate as many incoming requests as possible into the decoding process. Although it can always achieve high memory utilization, such an over-optimistic approach can lead to the peak memory required in the future by the running batch exceeding the hardware memory capacity, resulting in more frequent request evictions. In such cases, it results in response interruptions, particularly when deploying decode-heavy LLM services that require longer outputs, such as the popular ChatGPT o1-preview with reasoning capabilities.

To achieve optimal goodput for LLM services across varying request length distributions, we propose a novel Past-Future Scheduler. This scheduler takes into account both the historical output length distribution and the future required memory demand of the running batch, allowing it to schedule queued requests at optimal time points. It also strikes a balance between queue time and the frequency of request evictions, ultimately enhancing the performance of the serving system.

Specifically, the proposed scheduler consists of two components: predicting output length distribution and estimating future required memory. The output length distribution prediction component involves dynamically estimating the most likely final output length of requests within the running batch, based on the distribution of historical request output lengths and their current output lengths. It relies on the observation that actual output lengths of the requests are often less than the preset maximums, and the output distribution of historical requests remains stable within adjacent time windows. Whereas, the future-required memory estimation component involves dynamically calculating the memory usage required by the running batch at each future step after updating the output length prediction for each request. It considers the memory released by completed requests at future time points, thereby determining the future required memory needed for the entire current running batch. Therefore, the proposed scheduler can precisely assess the impact of adding a new request to the running batch on future required memory at different time steps. By balancing queuing time and the frequency of request eviction, it ensures requests are scheduled at the optimal time points. As shown in Figure \ref{fig:intro}, our method maintains high hardware memory utilization, with future required memory always approaching hardware capacity, while preventing evictions.

To demonstrate the effectiveness of the proposed Past-Future scheduler, we also designed a new LLM serving framework, LightLLM. With the help of the Past-Future scheduler, our LightLLM has been deployed to support services on a cluster with thousands of GPUs, providing various real-world LLM services such as code generation, human-like dialogue, long document analysis, and multimodal conversation. It can serve tens of millions of user requests and process billions of tokens per day under strict SLA constraints.

Specifically, this paper makes the following contributions:

\begin{itemize}
    \item We discover the similarity of the LLM requests output length distribution within adjacent time windows, and propose a method to predict the output length distribution based on recent historical requests.
    \item We propose an accurate estimation method for the future required memory of LLM requests, and combine it with our output length prediction method to propose the Past-Future Request Scheduler.
    \item We develop a high-performance LLM service framework, LightLLM, that implements the Past-Future scheduler in a real production system for serving tens of millions of user requests under strict SLA constraints.
    \item We evaluate LightLLM with comprehensive experiments to demonstrate our Past-Future scheduler can achieve better goodput performance on multiple hardware platforms.
\end{itemize}

\section{Background and Challenges}
\label{sec:background}
% 1 page
\subsection{LLM Inference and KV Cache}
Upon receiving a request consisting of a list of input prompt tokens, LLMs generate the output text in two stages: "prefill" and "decode". In the "prefill" stage, the model processes the input prompt tokens in parallel, producing an output token. Subsequently, the "decode" stage follows an autoregressive generation paradigm and generates output tokens one by one. Specifically, the input prompt tokens and the newly generated tokens are concatenated together and fed into the LLM to produce a new output token. This process is repeated iteratively until the end token is encountered or the maximum output length ($max\_new\_tokens$) is reached. However, inference with all tokens as input in each generation will lead to a lot of unnecessary computations, since in LLM the subsequent generated tokens will not affect the activation value of the previous token due to the presence of the attention mask. To speed up it, the Key and Value tensors of the input prompt tokens and the previously generated tokens are cached in GPU memory, known as KV cache. This caching mechanism allows the model to process only the latest input token when generating a new token, significantly enhancing token generation speed. With the KV cache, there is no need to recalculate the input prompt tokens and previously generated tokens during each decoding process. Only the newly generated tokens need to be fed into the LLM for the next step, reducing computational overhead and improving inference latency. However, while the KV cache can enhance inference speed, it also leads to significant GPU memory usage, making the system efficiency become memory-bound and limiting throughput performance. Thus, managing the KV cache has become a key challenge in deploying high-performance LLM services.

\subsection{Memory Management}

Current LLM services, such as FasterTransformer~\cite{FasterTransformer} and ORCA~\cite{yu2022orca}, typically store the KV cache in contiguous GPU memory space. When processing a request, they frequently allocate GPU memory based on the maximum potential output length, as the exact output length is hard to predict, leading to severe memory fragmentation and hampering the overall throughput. To address this, vLLM~\cite{kwon2023efficient} introduced PagedAttention, an attention algorithm that used block memory management, inspired by the classical virtual memory and paging techniques in operating systems. PageAttention reduces memory fragmentation by storing the KV cache in non-contiguous memory blocks. Despite the notable advances brought by memory management, current LLM frameworks underestimate the impact of the request scheduler on GPU memory utilization and system performance, and as a result, still fail to achieve optimal goodput under SLA constraints.

\subsection{Continuous Batching}
Batching multiple requests for inference is an effective way to improve the performance of LLM service. However, it presents greater challenges compared to traditional deep learning workloads, as it involves iteratively generating the final output text. If multiple requests with varying input lengths are statically grouped into a fixed batch size, padding them to match the length of the longest request leads to additional memory and computational waste. Moreover, due to the unpredictable output length of each request, earlier completed requests may be delayed while waiting for the unfinished ones. The continuous arrival of new requests may be queued and cannot be processed in a timely manner, resulting in a significant waste of both GPU memory and computational resources.

To address this issue, Continuous Batching, also known as Dynamic Batching or iteration-level scheduling, has been widely adopted in current LLM service frameworks. With customized attention operators and combined with memory management, when processing a batch of requests, there is no need to pad inputs to the maximum length or wait for all requests to finish. Instead, in each iteration, completed requests are filtered out promptly, and new requests are dynamically added to the batch based on the available GPU memory. As a result, individual requests can be processed and returned to the user more quickly, significantly reducing queuing time for new requests and improving overall system throughput.

For a typical implementation in popular frameworks, at a single decode iteration, the latest tokens of all current decoding requests are batched together for inference, and for the attention operation, a customized operator additionally accepts a list of the current request IDs and a mapping table (maintained by the memory management component) that associates these request IDs with their corresponding KV cache locations in the KV cache memory pool. The attention operator leverages this information to perform non-contiguous memory accesses within the KV cache memory pool, enabling it to compute attention operations for requests of varying sequence lengths. As for the linear operation in the LLM (which accounts for the vast majority of the model's weights), since it is independent among the tokens (the same weight is applied uniformly across all tokens), it can be performed directly as a single matrix multiplication for the batched tokens.

\subsection{Request Scheduler}

Current LLM frameworks control the continuous batching process through a request scheduler. The scheduler filters out completed requests based on the running state of the requests in the running batch and decides whether to add queued requests to the running batch based on the available system memory resources. Due to the unpredictable and highly varying output lengths of LLM requests, current LLM frameworks use either conservative or aggressive schedulers for request scheduling. A conservative scheduler estimates memory usage based on the sum of request input lengths and the $max_new_tokens$, which can significantly limit the number of requests processed simultaneously, thus reducing throughput. On the other hand, an aggressive scheduler, which disregards memory occupancy from request outputs and batches requests solely based on input lengths, may lead to excessive request evictions due to insufficient memory during inference. This can severely break SLA constraints and negatively impact the user experience.

\subsection{Challenges for Improving Goodput with SLA Guarantee}
\label{background:challenges}

A service-level agreement (SLA) is a contract between a service provider and its customers that documents what services the provider furnishes and defines the service standards the provider is obligated to meet~\cite{wiki:sla}. Systems that cannot meet the SLA guarantee lead to poor user experience. The key SLA metrics for LLM service include \textit{Time To First Token} (TTFT), \textit{Time Per Output Token} (TPOT), and \textit{Max Time Per Output Token} (MTPOT). TTFT measures the time it takes for users to receive the initially generated token. TPOT indicates the interval between each token received by users. MTPOT represents the maximum interval for users to receive each token, which is the maximum of TPOTs of a given request. Services that can guarantee SLA metrics for 99\% of requests can always be seen as stable and available. Throughput under SLA is considered effective throughput, also known as goodput.

Achieving high system goodput remains a significant challenge due to the difficulty of estimating how scheduling new requests affects the memory requirements of the running batch. First, calculating the memory demand of a running batch is inherently challenging. Simply summing the input lengths of all requests or combining the input lengths with the maximum output length is not suitable because of the dynamic nature of request arrivals and removals. Second, predicting the output length of requests is also difficult, necessitating a precise yet low-overhead estimation approach. This is why current frameworks often compromise by opting for either aggressive or conservative schedulers.

\section{Past-Future Request Scheduler}
\label{sec:method}

In this paper, we propose a novel Past-Future scheduler for LLM serving, which consists of the components of predicting output length distribution and estimating future required memory to achieve the precise scheduling of requests under SLA guarantees. The proposed scheduler can effectively address the challenges faced by existing request schedulers (discussed in Section~\ref{background:challenges})

\begin{figure*}[t]
  \centering
  \includegraphics[width=0.9\textwidth]{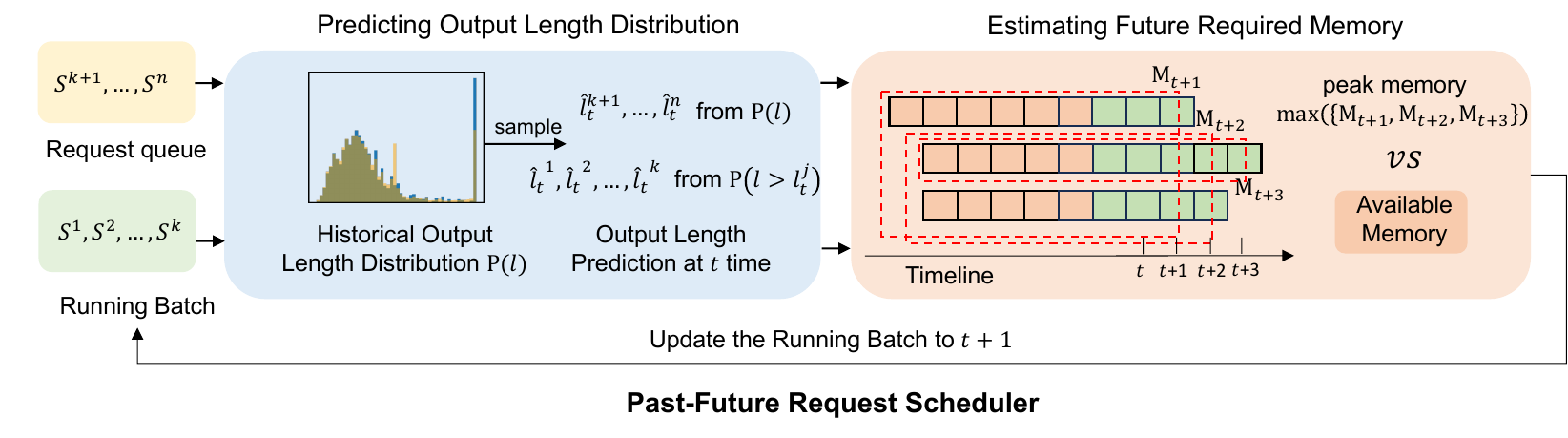}
  \caption{The overview of the proposed Past-Future scheduler. At time t, the service includes a running batch with $k$ requests and a request queue with $n-k$ requests. The proposed scheduler will first predict the most probable output length for queued requests and update the predicted output length of each request in the running batch based on the historical distribution of request output lengths and their output lengths at time $t$. Then it calculates the memory demand of the running batch at each future time if the requests are added at time $t$, and determines whether to schedule the requests at time $t$ based on whether any evictions would occur.}
  \label{fig:frame}
\end{figure*}

\subsection{Overview}
% 调度
As shown in Figure \ref{fig:frame}, the proposed Past-Future scheduler consists of two steps: predicting output length distribution and estimating future required memory. Specifically, at time $t$, the system has a running batch $B$ with $k$ requests $\{S^1, ..., S^k\}$ and a request queue $Q$ with $n -k$ waiting requests $\{S^{k+1}, ..., S^n\}$. The proposed Past-Future scheduler will first predict the output length distribution for waiting requests in $Q$ and update the predicted output lengths for requests in the running batch based on the distribution of historical request output lengths. Then, it will calculate the memory usage at each future time point after adding the waiting requests to the running batch, determine the future required memory, and schedule the requests based on the currently available memory in the system. With such an approach, it can flexibly schedule requests at the optimal time, achieving a reasonable trade-off between request queuing time and frequent request evictions, therefore maximizing system goodput. In the subsequent sections, we elaborate on how our scheduler predicts output length distribution (Section~\ref{subsec:predictingoutput}) and estimates future required memory (Section~\ref{subsec:estimatememory}).

\subsection{Predicting Output Length Distribution}
\label{subsec:predictingoutput}

To determine future memory requirements, we need to know the output length of the current requests. One approach is to use the maximum output length of the request, but since the LLM may end the output via EOS token at any time on its own, the actual output length may be much smaller than the maximum output length, leading to memory wastage. Therefore, accurately predicting the actual output length is essential for our scheduler.

However, predicting request output length is a highly challenging task. Existing prediction methods not only add extra inference overhead but also lack generalizability, making them impractical for real-world deployment. To address the issue, we first simplify the task to predict the output length distribution for all current requests, instead of seeking to accurately predict individual requests. After that, we can improve the prediction accuracy for individual requests by updating the prediction for each request on the fly based on the request's already decoded length at runtime, allowing it to work with our module to estimate future required memory (Section~\ref{subsec:estimatememory}) in order to achieve the high memory utilization scheduling without frequent request evictions.

For the output length distribution prediction, we analyzed multiple trace datasets from online LLM services including chat, code completion, and general API service, which leads to an interesting observation: \textit{the distribution of request output lengths within a certain window size is consistently stable for some services (usually a single service or task type), whereas the distribution of output lengths for other hybrid task or API services may vary over a long time period (hours or days), but still tends to be stable within a short time period (minutes)}. This means that both types of service requests in two adjacent sliding time windows are likely to have similar output length distributions. Therefore, we can effectively estimate the output length for requests of the currently running batch based on the historical distribution of recently finished requests.

\begin{figure*}[t]
  \centering
  \includegraphics[width=\textwidth]{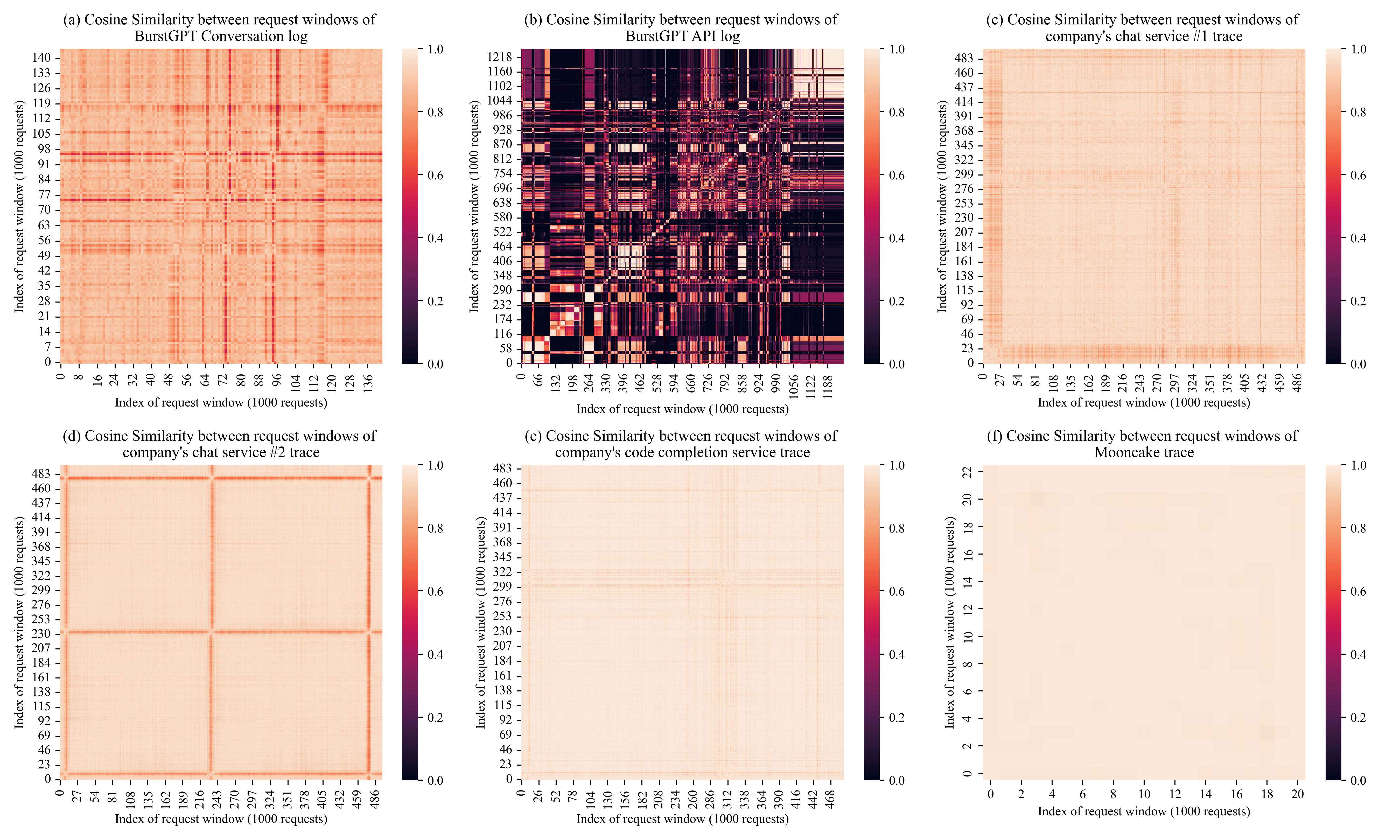}
  \caption{Cosine similarity of output length distributions (higher values indicate higher similarity) between the different partitioned time windows (1000 requests, no overlap) of multiple request trace dataset, points indicating the overlapped request window are excluded. The adjacent time windows (the diagonal pattern) have similar distributions (some dataset has similar distributions globally).}
  \label{fig:dist}
\end{figure*}

\begin{figure}[t]
  \centering
  \includegraphics[width=\linewidth]{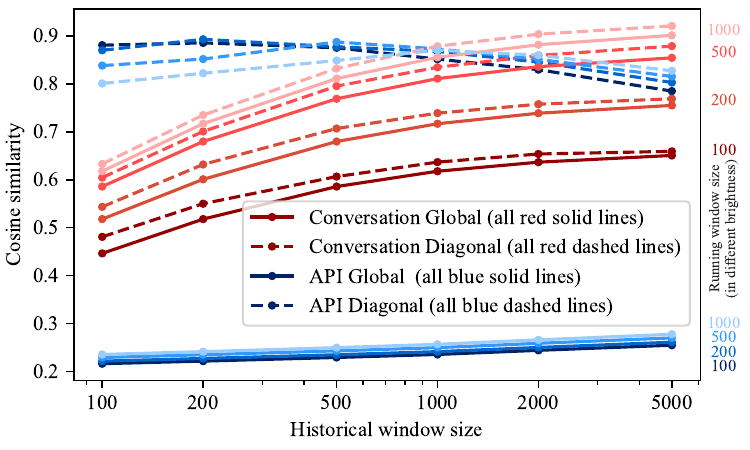}
  \caption{Average cosine similarity on global or diagonal line under different historical (x-axis) and running request windows size (in different brightnesses of a color, corresponding to the number markers on the right) in the BurstGPT dataset.}
  \label{fig:cos_win}
\end{figure}

Figure~\ref{fig:dist} shows the cosine similarity between the output length distribution at different time windows on multiple online service requests trace datasets. BurstGPT~\cite{wang2024efficient} contains trace data for over 1,400,000 requests made to ChatGPT and GPT-4. The requests are divided into two categories: (a) online conversation service requests, and (b) API service requests, which are the two most common invocation methods for LLM services. We have also validated our approach on more datasets from our in-house services (c, d, e) and the Mooncake~\cite{qin2024mooncake} trace from Moonshot (f). We find that for a single end-user service, including dialog (a, c, d, f) or code completion (e), its overall distribution is less varying. For API services (b), on the other hand, its overall distribution is more varying as it usually mixes multiple service types (to the best of our knowledge, there are no more publicly available trace datasets for LLM API services or hybrid tasks yet). However, for all datasets, the adjacent time windows (the diagonal pattern) have similar distributions. We believe that the services in the above datasets cover the vast majority of current scenarios of LLM service, and thus this distribution similarity of adjacent time windows is universal.

In practice, the size of the historical request window (the preceding window in adjacent windows) can be freely adjusted, whereas the size of the current request window (the succeeding window in adjacent windows) is determined by the current batch size. To evaluate the effectiveness under various workload conditions, we conducted experiments with different combinations of window sizes on the BurstGPT dataset. The conversation logs from the BurstGPT dataset are shown in red, and the API logs are shown in blue, with different brightnesses of a color representing various running window sizes, as illustrated in Figure~\ref{fig:cos_win}. The results show the distribution similarity of adjacent windows (diagonal) is maintained at a high level across multiple running window sizes. Variations in the historical window size exhibit distinct trends for different request types, but a size of 1,000 balances both types effectively. For simplicity, we have adopted this setting uniformly in subsequent experiments, as further tuning yields negligible improvements to scheduling effectiveness.

Based on the above observations, we introduce an efficient, parameter-free prediction method, for predicting output length distribution, which estimates and dynamically updates the output lengths for each request in the current running batch at each time step based on the historical distribution of recently finished request output lengths. 

Specifically, the proposed method utilizes the probability distribution of historical request output lengths to dynamically predict the output length of current requests at each inference step. During serving, it records the actual output lengths of historical requests from adjacent time windows, denoted $L_{h} = \{l^0_{h}, l_h^1, ..., l_h^w\}$ where $w$ is the window size. Thus the probability distribution can be described as:
\begin{equation}
    P(l) =  \mathcal{C}(l, L_h) / w
    \label{eq:5}
\end{equation}
where $\mathcal{C}(l, L_h)$ is a counter function, which returns the number of elements in $L_h$ equal to $l$.

Therefore, as shown in Figure \ref{fig:frame}, at time step $t$, before {estimating future required memory, the proposed scheduler will first sample a $\hat{l}_t^i$ for each request $S^i$ in the request queue $S_q = \{S^{k+1}, ..., S^n\}$ from $P(l)$ and update the predicted length $L_{t-1} = \{\hat{l}_{t-1}^1, \hat{l}_{t-1}^2,$\newline $ ..., \hat{l}_{t-1}^k\}$ by resampling $\hat{l}_t^j$ for each request $S^j$ in the running batch $S = \{S^0, S^1,..., S^n\}$ from $P(l > l_{t-1}^j)$ to get a new predicted output length $L_{t} = \{\hat{l}_{t}^1, \hat{l}_{t}^2, ..., \hat{l}_{t}^k\}$.

By dynamically updating the output length of the request at each step, the proposed scheduler can ensure that the predicted values are always approaching the real output length distribution and the predicted length is larger than the actual output length as much as possible. Therefore, it can not only accommodate as many requests as possible, but also ensures sufficient memory to guarantee the completion of requests, thereby avoiding request eviction. Additionally, in contrast to some existing methods that obtain the output length by classifying the input content or fine-tuning the LLM, our method has very high generalizability (model-independent) and without incurring any additional inference overhead. Algorithm~\ref{alg:sche} describes the details of our proposed scheduler to handle new requests at time $t$.

\begin{algorithm}
    \caption{Past-Future Scheduler}
    \label{alg:sche}
    \begin{algorithmic}[1]
        \Statex \textbf{Input:} historical output lengths $L_h$, request queue $S_q$, running batch $S$ with the number of generated tokens $\{l_t^1, l_t^2, ..., l_t^k\}$ at time $t-1$, system memory capacity $\mathbf{M}$.
        \Statex \textbf{Output:} the updated running batch $S$ 
        \State Build the probability distribution of output length $P(l)$ with $L_h$, via Equation \ref{eq:5}
        \State Initialize the predicted output length $L_t = \{\}$
        \For{$l_t^i$ in $\{l_t^1, l_t^2, ..., l_t^k\}$}
            \State Sample $\hat{l}_t^i$ from $P(l>l_t^i)$
            % \State $L_t$ $\longleftarrow$ $\hat{l}_t^i$
            \State $L_t \gets L_t \cup \{\hat{l}_t^i\}$
        \EndFor
        \For{$S^j$ in $S_q$}
            \State Sample $\hat{l}_{t}^j$ from $P(l)$
            % \State $L_t$ $\longleftarrow$ $\hat{l}_{t}^j$
            \State $L_t \gets L_t \cup \{\hat{l}_{t}^j\}$
            \State Calculate future required memory $\mathbf{M}_t^*$ with $L_t$ (and other known information), via Equation \ref{eq:1} to \ref{eq:3}
            \If{$\mathbf{M}_t^* <= \mathbf{M}$}
                % \State  $S$ $\longleftarrow$ $S^j$
                \State $S \gets S \cup \{S^j\}$
                % \State $L_t \gets L_t \cup \{\hat{l}_{t}^j\}$
            \Else
                \State Return $S$
            \EndIf
        \EndFor
    \end{algorithmic}
\end{algorithm}

\subsection{Estimating Future Required Memory}
\label{subsec:estimatememory}

\begin{figure}[t]
  \centering
  \includegraphics[width=0.28\textwidth]{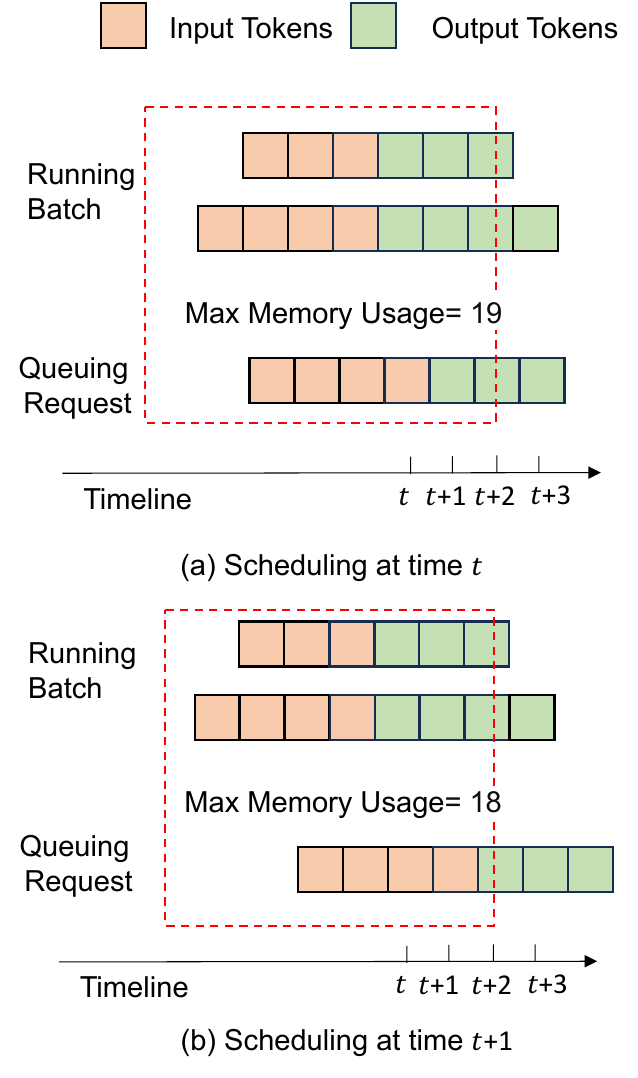}
  \caption{Memory demand of a running batch when the queuing request is scheduled at different time steps.}
  \label{fig:peak_1}
\end{figure}

After predicting the output length of each request, the memory demand of the running batch can be calculated.
Estimating the memory demand for the entire running batch remains challenging due to the varying entry and completion times of requests. Current schedulers, which statically estimate memory usage, can only account for memory release when a request is completed. This results in either low memory utilization with a conservative approach or frequent request evictions with an aggressive approach. As shown in Figure \ref{fig:peak_1}, even if no request is completed, adding the same request into the running batch at different time steps leads to varying memory demands. Therefore, our scheduler incorporates a novel method to consider the future process when scheduling new queued requests. By anticipating the memory resources released by future completed requests, it dynamically schedules queued requests at optimal time points, minimizing queuing delays while preventing frequent evictions during inference. This ensures high goodput while maintaining SLA guarantees.

\begin{figure*}[t]
  \centering
  \includegraphics[width=0.85\textwidth]{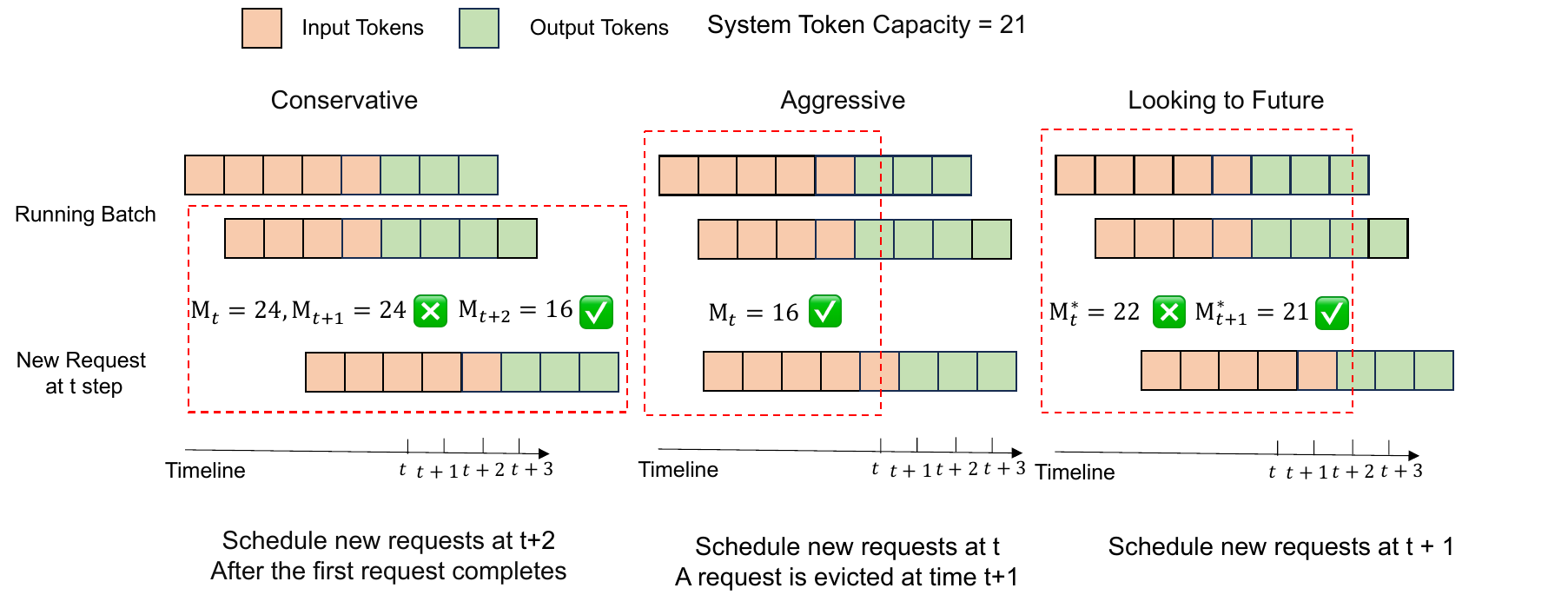}  % 将 "your_image_file" 替换为实际的图像文件名
  \caption{Behavior comparison of different schedulers when handling new requests at time t. It shows that our method can introduce requests at the appropriate time, maintaining high memory utilization while avoiding harmful request evictions.}
  \label{fig:peak}  % 将 "your_label" 替换为图像的标签，用于引用
\end{figure*}

Specifically, given a running batch $S = \{S^1, S^2, ..., S^k\}$ with a predicted output length $L_t = \{\hat{l}_t^1, \hat{l}_t^2, ..., \hat{l}_t^k\}$ at step $t$, we can accurately calculate the memory requirement for inference of each request, which is the sum of the input length and output length. However, the memory required to complete the entire running batch of requests is not simply the sum of the memory needed for all individual requests, since each request in $S$ will finish at different time steps and dynamically release the memory resource. Simply using the sum of the memory of all requests to determine whether the system has sufficient memory to complete the currently running batch successfully leads to low memory utilization. In contrast, the required memory at future time points is the minimum memory requirement to complete the current batch.

As illustrated on the right side of Figure \ref{fig:frame}, we note that the future required memory will inevitably occur at the moment a request ends. If no request ends at time point $t$, the GPU memory usage of the running batch at $t$ must be less than that at $t+1$. At the $t+2$ time point, the request with the shortest output completes and releases the memory it occupies, reducing the memory occupancy of service to $\mathbf{M}_{t+2}$. At $t+3$ time, this leaves only one request that needs to be processed, which requires the memory of $\mathbf{M}_{t+3}$.
Therefore, only $max(\mathbf{M}_{t+1}, \mathbf{M}_{t+2}, \mathbf{M}_{t+3})$ space is needed to complete the inference of requests in the batch. To calculate memory usage when each request is completed, the requests are first sorted in descending order based on the estimated remaining generation length.

\begin{equation}
\begin{aligned}
& \quad \quad \quad \{(l_p^i + l_t^i, \hat{l}_t^{i} - l_t^i) | S^i \in S\} \\
& where \quad  \hat{l}_t^{i} - l_t^i \geq \hat{l}_t^{i+1} - l_t^{i+1}, i=1,2,..., k
\end{aligned}
\label{eq:1}
\end{equation}
where $l_p^i$ is the input length of request $S^i$, $l_t^i$ is the number of generated tokens of request $S^i$ at time t, $\hat{l}_t^{i}$ is the predicted output length of request $S^i$, $k$ is the current batch size. Then, we can calculate the memory occupancy of the system when the $S^i$ request finishes:
\begin{equation}
\begin{split}
    \mathbf{M}_i =& (\sum_{j=1}^i{\{l_p^j + l_t^j\}}) + (\hat{l}_t^{i} - l_t^i) * i
\end{split}
\label{eq:2}
\end{equation}

Finally, we can determine that the future required memory to complete Batch $S$ is:
\begin{equation}
    \mathbf{M}^* = max(\{\mathbf{M}_i|i=1,2,...,k\})
    \label{eq:3}
\end{equation}

Figure \ref{fig:peak} illustrates the differences between the proposed scheduler and the conservative or aggressive schedulers currently used by existing frameworks when handling new requests. Assume that the system has a total capacity of 21 tokens. The conservative scheduler only adds new requests when it is ensured that the system memory resources exceed the total resources required by all requests (at time $t$+2, the first request has been completed). It not only results in high response latency, but also leads to low utilization of hardware resources. The aggressive scheduler, on the other hand, immediately adds a request if it finds sufficient system memory to handle the input (at time $t$). However, as inference progresses, it often discovers that the system resources are insufficient (at time $t$+2, $M_{t+2}=22 > 21$), necessitating the eviction of a request to ensure that inference can continue (which requires request re-queuing and recomputation). In contrast, our scheduler evaluates the impact of adding requests at different times on the future required memory usage of the system. Consequently, it schedules requests at the most appropriate time (at $t+1$), allowing requests to wait for a few decoding steps to avoid eviction. It ensures better utilization of system resources while reducing the risk of request evictions.

\section{Implementation}
To demonstrate the effectiveness of the proposed Past-Future scheduler, we developed a high-performance LLM serving framework, LightLLM. LightLLM is a Python-based framework for LLM inference and serving, designed to be lightweight, easily scalable, and high-performance. It realizes multi-process asynchronous collaboration, parallelizing request pre-processing, scheduling with model inference, and post-processing. The front end of LightLLM is built using FastAPI, while the backend inference relies primarily on PyTorch~\cite{paszke2017automatic} and OpenAI Triton~\cite{openai_triton}. A wide range of models has already been supported, including BLOOM~\cite{workshop2023bloom}, Llama~\cite{touvron2023llama}, Llama2~\cite{touvron2023llama}, StarCoder~\cite{li2023starcoder}, ChatGLM2-6b~\cite{du2022glm,zeng2022glm}, Qwen~\cite{qwen}, Baichuan~\cite{baichuan2023baichuan2}, Baichuan2~\cite{baichuan2023baichuan2}, InternLM~\cite{2023internlm}, and Yi~\cite{2023Yi}. Moreover, LightLLM also supports the deployment of multimodal large language models, such as LLaVA~\cite{liu2023llava} and Qwen-VL-Chat~\cite{Qwen-VL}. The complete code has been \href{https://github.com/ModelTC/lightllm}{open-sourced}.

\noindent\textbf{Scheduler workflow.} We implemented the Past-Future scheduler using parallel computation on GPUs in LightLLM, which has a low overhead (less than 1\% of LLM model inference time). We evaluated different historical request window sizes from hundreds to thousands on multiple datasets, which all worked well with our scheduler. Therefore, we chose 1,000 as an efficient and widely adopted setting. When the size of the running batch is low, the sampling prediction is repeated several times to improve accuracy. Moreover, at the initial stage of service startup, we initialize the output length distribution using the preset maximum output length, which can be updated quickly in a few minutes.

\section{Evaluation}
\label{sec:evaluation}
In this section, we provide a comprehensive evaluation of our framework across multiple models and various hardware platforms to prove its superiority over state-of-the-art LLM request schedulers.

\subsection{Experimental Setup}
\noindent\textbf{Model.} To demonstrate the effectiveness of our scheduler, we choose Llama2~\cite{touvron2023llama2}, the most commonly used model in both industry and academia, for performance analysis. We evaluate models of three different scales, 7B, 13B, and 70B, to compare performance across different schedulers implemented in LightLLM. Additionally, we evaluate the performance improvement of our scheduler on multimodal tasks using popular models such as Qwen-VL-Chat~\cite{Qwen-VL} and LLaVA-1.5~\cite{liu2023improved}.

\noindent\textbf{Dataset.} 
To evaluate the performance of different schedulers under varying output length distributions, we experiment three datasets with different distributions (\textit{Distribution-1, -2, -3}) based on the request distribution of our real-world LLM service, with \textit{input/output} length uniform distributions of \textit{32-4k/2k-4k}, \textit{3k-5k/3k-5k}, and \textit{2k-4k/32-4k}. We also experiment with the ShareGPT-o1 dataset using the ShareGPT dataset and the OpenAI o1-preview API. By sending user questions from the ShareGPT dataset to the API, we collect output token counts, creating a request-length dataset reflective of the Chain of Thought model serving workloads. Additionally, for multimodal performance assessment, we adopt the validation set of the TextVQA~\cite{singh2019towards} dataset, which contains 5,000 questions and 3,166 images.

\noindent\textbf{SLA guarantee.} 
A service that meets SLA guarantees is a usable service. Therefore, we mainly focus on the throughput comparison of LLM service frameworks equipped with different schedulers under SLA guarantees. Therefore, we only measure the throughput of requests that meet the SLA, which we refer to as goodput.

Common LLM applications such as chatbots use streaming output methods, where Time To First Token (TTFT) is the time from when the user sends a request to when the first output token is received, and Time Per Output Token (TPOT) is the delay between each subsequent output token. For the average TPOT, it is generally accepted that it needs to be larger than the human reading speed, which is usually satisfied by existing frameworks. Equally important is the maximum TPOT in a request (MTPOT), which is one of the metrics we use subsequently. This is because as soon as there exists one high TPOT, the user will observe a noticeable output stall, while the average TPOT or even the P99 TPOT can hardly reflect this occurrence.

In the following experiments, we define the SLA guarantee based on the hardware capabilities and model sizes: ($TTFT < 10$ seconds, $MTPOT < 1.5$ seconds) for 7B/13B models, and ($TTFT < 15$ seconds, $MTPOT < 5$ seconds) for 70B model across all hardware.

\begin{figure*}[t]
  \centering
  \includegraphics[width=0.96\textwidth]{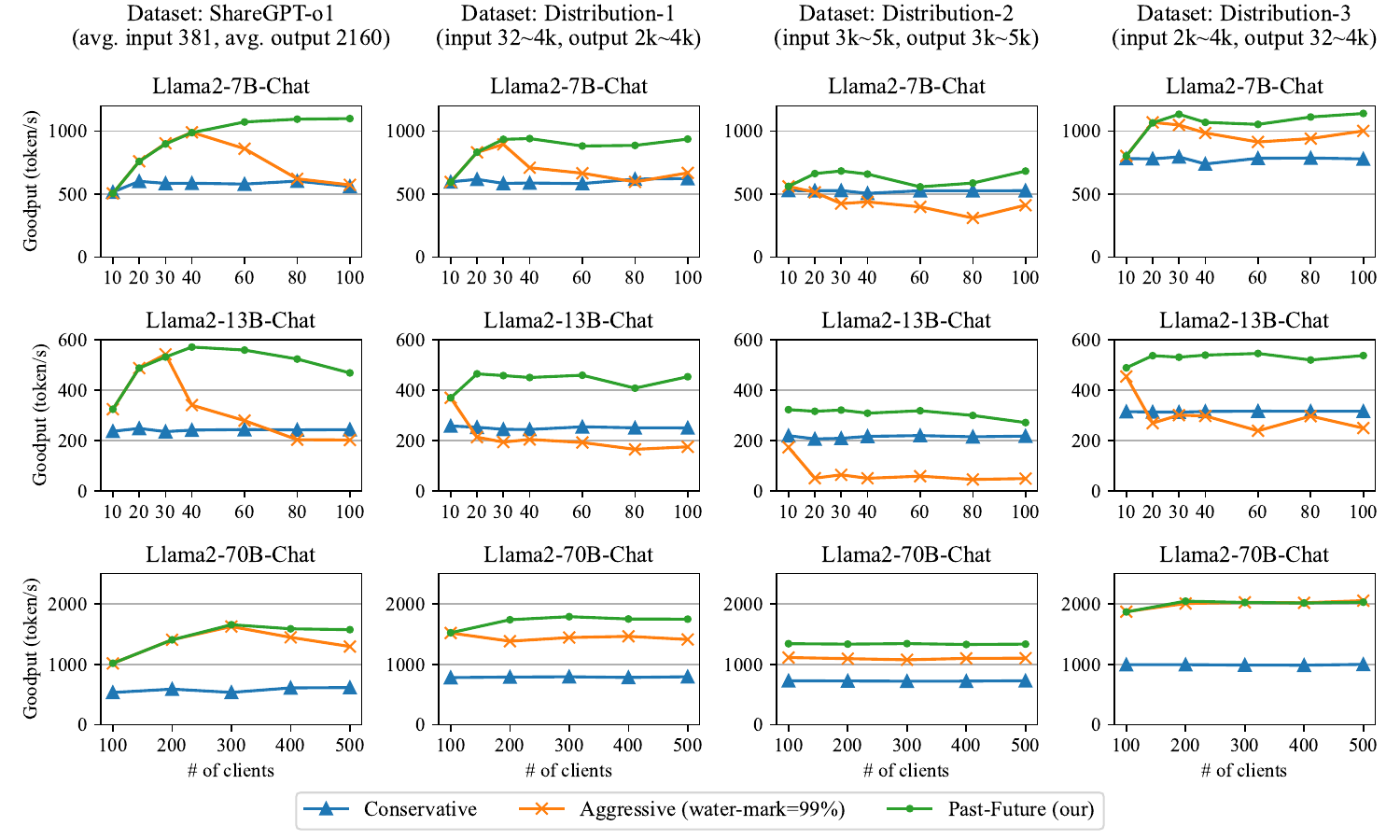}
  \caption{Detailed goodput under different numbers of clients for different schedulers. Conservative schedulers can only achieve low goodput at all times, while aggressive schedulers, although they can achieve equally goodput under light loads, have a rapid degradation of goodput under high loads. Our Past-Future scheduler maintains a consistently high goodput.}
  \label{fig:sla}
\end{figure*}
\subsection{Comparison with Different Schedulers}
Figure \ref{fig:sla} shows the goodput of different schedulers across various model sizes and datasets on NVIDIA A100-80G (1 GPU for 7B and 13B models, 4 GPUs with NVLink for 70B model). It can be observed that when there are few concurrent clients, the system memory load is low, resulting in the same goodput performance across different schedulers. As concurrency increases, conservative schedulers waste a significant amount of memory, causing many requests to be queued and leading to first-token latency breaking the $TTFT$ SLA limit, resulting in lower goodput compared to aggressive schedulers and the proposed Past-Future scheduler. As concurrency continues to rise to full system load, aggressive schedulers suffer from repeated evictions, causing requests to violate the $MTPOT$ SLA constraint, which leads to a decline in goodput. In contrast, our Past-Future scheduler accurately estimates the maximum memory demand of each running batch, ensuring that all scheduled requests are completed successfully. Thus, as the number of concurrent clients increases, the goodput eventually stabilizes, reaching the maximum that can be handled by the hardware under the given SLA constraints.

Moreover, it is worth noting that the schedulers exhibit different performances under varying input-output length distributions. In prefill-heavy datasets (where input length > output length), such as \textit{Distribution-3}, both the aggressive scheduler and our Past-Future scheduler achieve strong performance, surpassing the conservative scheduler. However, in decode-heavy datasets (where output length > input length), such as \textit{ShareGPT-o1} and \textit{Distribution-1}, the aggressive scheduler, due to ignoring output memory requirements, leads to frequent eviction of requests, causing goodput to first increase and then decline. In contrast, our proposed scheduler ensures stable performance across various LLM scenarios, regardless of the input-output length distribution.

\begin{table*}[t]
\centering
\small
\caption{Performance of different scheduling methods on multiple datasets.}
\label{tab:sche_acc}
\begin{tabular}{clrrrr}
    \toprule
    \multirow{2}{*}{Dataset} & \multirow{2}{*}{Method} & \multirow{2}{*}{Decoding Steps} & Current Consumed      & Future Required       & \multirow{2}{*}{Evicted Reqs} \\
                             &                         &                                 & Memory & Memory     & \\
    \midrule
    \multirow{3}{*}{\shortstack{Distribution-1 \\ (Decode-heavy)}}
    & Theoretical optimum                   & 294250       & 94.87\%     & 98.89\%         &  0\% \\
    & \textbf{Past-Future (reserved=3\%)}   & \textbf{296710}       & \textbf{93.61\%}     & \textbf{97.47\%}         &  \textbf{6.86\%} \\
    & \textbf{Past-Future (reserved=5\%)}   & \textbf{301680}       & \textbf{91.87\%}     & \textbf{95.73\%}         &  \textbf{3.37\%} \\
    & \textbf{Past-Future (reserved=10\%)}  & \textbf{320720}       & \textbf{87.09\%}     & \textbf{91.51\%}         &  \textbf{1.58\%} \\
    & Aggressive (watermark=99\%)           & 284160       & 98.41\%     & 103.12\%        &  93.74\% \\
    & Aggressive (watermark=95\%)           & 290980       & 96.10\%     & 101.21\%        &  24.08\% \\
    & Aggressive (watermark=90\%)           & 302900       & 92.46\%     & 97.87\%         &  10.76\% \\
    & Conservative (no overcommit)          & 485650      & 57.75\%     & 60.79\%         &  0\% \\
    & Conservative (overcommit=150\%)       & 320530      & 87.34\%     & 90.79\%         &  17.23\% \\
    \midrule
    \multirow{3}{*}{\shortstack{Distribution-2 \\ (Balanced)}}
    & Theoretical optimum                   & 653120       & 92.57\%     & 98.29\%         &  0\% \\
    & \textbf{Past-Future (reserved=3\%)}   & \textbf{654500}       & \textbf{92.16\%}     & \textbf{97.21\%}         &  \textbf{7.42\%} \\
    & \textbf{Past-Future (reserved=5\%)}   & \textbf{669770}       & \textbf{90.07\%}     & \textbf{95.82\%}         &  \textbf{4.39\%} \\
    & \textbf{Past-Future (reserved=10\%)}  & \textbf{705920}       & \textbf{85.26\%}     & \textbf{90.99\%}         &  \textbf{1.54\%} \\
    & Aggressive (watermark=99\%)           & 616230       & 97.83\%     & 103.68\%        &  97.72\% \\
    & Aggressive (watermark=95\%)           & 628410       & 95.86\%     & 101.81\%        &  26.19\% \\
    & Aggressive (watermark=90\%)           & 654190       & 92.37\%     & 99.12\%         &  13.88\% \\
    & Conservative (no overcommit)          & 842470      & 71.71\%     & 75.45\%         &  0\% \\
    & Conservative (overcommit=125\%)       & 665970      & 90.64\%     & 94.68\%         &  84.34\% \\
    \midrule
    \multirow{3}{*}{\shortstack{Distribution-3 \\ (Prefill-heavy)}}
    & Theoretical optimum                   & 230690       & 96.60\%     & 98.60\%         &  0\% \\
    & \textbf{Past-Future (reserved=3\%)}   & \textbf{239260}       & \textbf{94.50\%}     & \textbf{96.51\%}         &  \textbf{2.59\%} \\
    & \textbf{Past-Future (reserved=5\%)}   & \textbf{241650}       & \textbf{92.64\%}     & \textbf{94.62\%}         &  \textbf{0.87\%} \\
    & \textbf{Past-Future (reserved=10\%)}  & \textbf{263670}       & \textbf{87.80\%}     & \textbf{89.87\%}         &  \textbf{0.00\%} \\
    & Aggressive (watermark=99\%)           & 227250       & 98.13\%     & 100.20\%        &  34.30\% \\
    & Aggressive (watermark=95\%)           & 234540       & 95.07\%     & 97.16\%        &  3.23\% \\
    & Aggressive (watermark=90\%)           & 247130       & 90.29\%     & 92.33\%         &  0.17\% \\
    & Conservative (no overcommit)          & 374250      & 59.73\%     & 61.59\%         &  0\% \\
    & Conservative (overcommit=150\%)       & 246870      & 90.45\%     & 92.54\%         &  19.09\% \\
    \bottomrule
\end{tabular}
\end{table*}

\subsection{Ablation Study}
In this part, we conduct an in-depth analysis of the Past-Future scheduler on decode-heavy \textit{Distribution-1}, balanced \textit{Distribution-2} and prefill-heavy \textit{Distribution-3} from two perspectives: memory utilization and request eviction. Table~\ref{tab:sche_acc} lists the decoding steps, average memory utilization, average future required memory, and the percentage of evicted requests across different schedulers of Llama-2 7B Chat on NVIDIA A100 80G. We also added the overcommit support for Conservative (e.g. overcommit=150\% means that the scheduler assumes 1.5 times the actual memory capacity, we reduce the overcommit ratio for the Conservative scheduler in Distribution-2 due to too many evictions).

In Table~\ref{tab:sche_acc}, the Decoding Steps represent the number of decodes required by continuous batching to perform these requests, a higher value means that these requests are being performed with a smaller batch size per decode and more times of decoding, which negatively affects the throughput performance. Current Consumed Memory is the average memory utilization during runtime, and the Future Required Memory is the actual memory requirement in the future to finish all current added requests when scheduling, more than 100\% means that the new requests added at this time will lead to a lack of memory in the future, the table shows the average value. Evicted Reqs is the ratio of the number of request evictions to the total number of requests. Request evictions not only violate the SLA of the requests, but also lead to a reduction in throughput performance due to the need for recomputation or swapping.

We also performed a detailed evaluation of various scheduler parameter configurations. We constructed the test dataset by concatenating the ShareGPT-01 dataset followed by the Distribution-1, Distribution-2, and Distribution-3 in sequence to generate a workload with varying output length distributions. The results shown in Figure~\ref{fig:scheduler_adv} demonstrate that the Past-Future scheduler outperforms both the aggressive and conservative schedulers under all configurations.

\begin{figure}[t]
  \centering
  \includegraphics[width=\linewidth]{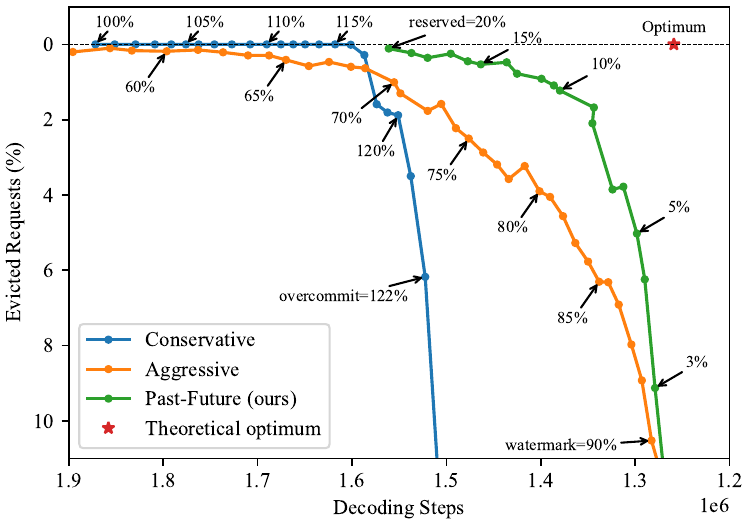}
  \caption{Comparison of different scheduler parameters on a varying load consisting of ShareGPT-o1 and Distribution-1 to 3.}
  \label{fig:scheduler_adv}
\end{figure}

\noindent\textbf{Effect on Memory Utilization.} 
It can be seen that the conservative scheduler, assuming that each request is expected to generate $max\_new\_token$ outputs, results in the lowest memory utilization and takes the most decoding steps. 
The aggressive scheduler, which does not consider the memory occupancy of request outputs, processes as many requests as possible simultaneously until achieving a given memory watermark, it can achieve high memory utilization. In contrast, the proposed Past-Future scheduler which estimates the memory occupancy of each request at each future time step based on the dynamically predicted output lengths sampled from the probability distribution of historical requests, can achieve high GPU memory utilization close to the theoretical optimum on all datasets. The theoretical optimum represents the optimal case where the memory is optimally utilized when the request output length is known, which is impossible in real-world service.

\noindent\textbf{Effect on Request Eviction.} 
Although an aggressive scheduler with a high memory watermark or a conservative scheduler with overcommit can also achieve high memory utilization, it introduces a significant amount of request eviction as shown in Table \ref{tab:sche_acc} (93.74\% on decode-heavy \textit{Distribution-1} for aggressive scheduler with watermark=99\%). Tuning the watermark or overcommit ratio helps to reduce evicted requests, but decoding steps also increase. At the same time, this tuning is more difficult to adapt to the workload with variable distribution, we further analyze the different parameters of different schedulers in this case, as shown in Figure~\ref{fig:scheduler_adv}, both aggressive and conservative have difficulty in balancing the evicted requests and decoding steps, and cannot achieve the effect of Past-Future scheduler by parameter tuning. Since aggressive and conservative schedulers are not able to sense the actual load, their evicted requests increase rapidly when approaching the memory capacity threshold. The superior performance of the Future-Past scheduler arises from its ability to sense the workload through historical distribution, enabling it to accurately predict future memory requirements. Although it incorporates a percentage of reserved memory in its scheduling decisions to handle minor fluctuations in request distribution, it consistently outperforms other schedulers across all reserved ratio configurations, and the eviction rate exhibits smoother variation within an appropriate range.

\subsection{End-to-end Performance}

To demonstrate that our improvements were not achieved based on a lower baseline framework, we compare LightLLM with the proposed Past-Future scheduler with the currently popular LLM service frameworks, including Text-Generation-Inference (TGI)~\cite{tgi} with conservative scheduler, vLLM~\cite{kwon2023efficient} with aggressive scheduler, and DeepSpeed-MII (FastGen)~\cite{fastgen} with conservative scheduler and the splitfuse strategy of the recent version on the commonly-used ShareGPT dataset. To simulate the unknown output length in a real-world scenario, we set a maximum output length $max\_new\_tokens$ $=$ 2048 for each request. Requests will be completed when they reach the maximum length or encounter an end token. We also implemented a conservative scheduler for the TensorRT-LLM~\cite{trtllm} backend.

We evaluate the maximum throughput and the goodput under the SLA guarantee of each framework which adopts different scheduling strategies by simulating concurrent requests from different numbers of clients. Figure \ref{fig:e2e} presents the performance across different hardware platforms and model sizes. It can be observed that with the help of the proposed Past-Future scheduler, LightLLM performed a competitive performance on maximum throughput, and a superior goodput in all scenarios. Additionally, we find that TGI and DeepSpeed-MII, employing a conservative scheduling approach, fail to achieve high maximum throughput, suggesting that a conservative scheduling strategy can lead to the wastage of computational resources and memory resources. Even with the fast static inference speed of TensorRT-LLM, it is difficult to achieve satisfactory throughput with a conservative scheduler.

In contrast, vLLM, which adopts an aggressive scheduler, achieves better maximum throughput performance with 7B and 13B models. However, due to the aggressive scheduler, which can more easily introduce harmful request eviction and re-queue, there is a noticeable decline in goodput under the SLA guarantee. Equipped with the Past-Future Scheduler, LightLLM can maximize its maximum throughput while avoiding harmful request eviction, thus ensuring goodput under the SLA guarantee. \footnote{The performance data for each framework, including LightLLM, in Figure \ref{fig:e2e} is based on the version from December 2023. The comparison across frameworks is only intended to show that our scheduler achieves improvements based on a strong baseline. Each framework has seen significant progress and development since then, and we will release the updated results in our open-source code repository.}

\begin{figure}[t]
  \centering
  \vspace{-3pt}
  \includegraphics[width=0.95\linewidth]{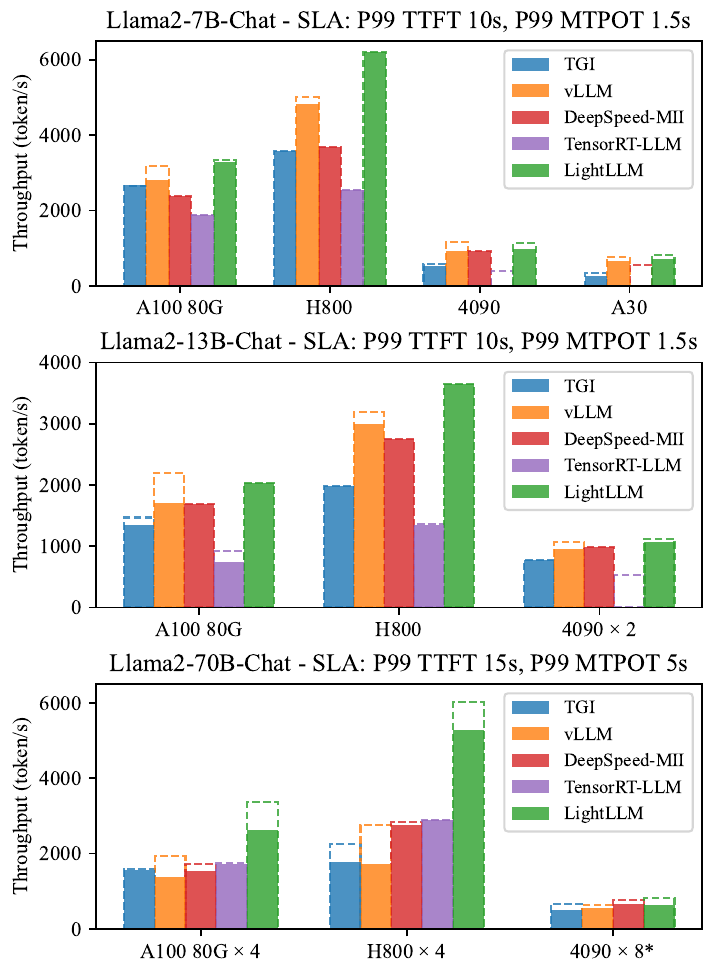}
  \caption{Throughput (dashed) and SLA-guarantee goodput (solid) of different frameworks across various hardware. SLA constraints marked as \textit{P99 TTFT 10s, P99 MTPOT 1.5s} refers to the 99\% of the requests' first token latency must be < 10s, and the requests' maximum per-token latency for 99\% requests must be < 1.5s.}
  \label{fig:e2e}
\end{figure}

\subsection{Performance on Multimodal LLM}
We also validate the compatibility of the proposed Past-Future scheduler on multimodal tasks. Table \ref{tab:multi} presents the throughput comparison between the original implementations and LightLLM with the Past-Future scheduler for Qwen-VL-Chat, LLaVA-1.5-7B, and LLaVA-1.5-13B. Due to the Past-Future scheduler's ability to effectively improve system memory utilization, we achieved a 50\% throughput improvement on Qwen-VL-Chat and a 60\% improvement on LLaVA-1.5-7B compared to their original implementation.

\begin{table}[h]
\centering
\small
\caption{Performance comparison of LightLLM on the Qwen-VL-Chat, LLaVA-1.5-7B and LLaVA-1.5-13B with their original implementation.}
\label{tab:multi}
\begin{tabular}{lrr}
\toprule
& \multicolumn{2}{c}{Throughput (tokens/s)} \\
\cmidrule(lr){2-3}
Model & Origin & \textbf{LightLLM} \\ 
\midrule
Qwen-VL-Chat & 219.96 & \textbf{319.19} \\
Llava-1.5-7B & 535.31 & \textbf{851.73} \\
Llava-1.5-13B & 1193.38 & \textbf{2228.71} \\
\bottomrule
\end{tabular}
\end{table}

\section{Related Work}
\label{sec:related_work}

With the popularity of large-scale model applications, there is an increasing demand for deploying high-performance large models. Currently, there are some excellent frameworks for serving large models. They optimize the throughput performance of large model services from multiple perspectives, including parallelization, GPU operator optimization, request batching strategies, and GPU memory management. 

Parallelization is the most commonly used technique to accelerate the LLM inference which includes data parallelism, pipeline parallelism, tensor parallelism, and sequence parallelism. Data parallelism and sequence parallelism are often employed for accelerating LLM training. Pipeline parallelism and tensor parallelism are widely applied to large model inference. Pipeline parallelism splits the model into multiple GPUs, with each GPU loading one or more layers of the LLM. While tensor parallelism involves splitting the model's parameters (tensors) across different devices (such as GPUs) and performing computations in parallel. Both of them can make the system handle the substantial parameter sizes of these models and achieve superior overall performance.

GPU operator optimization is also an effective way to speed up the inference because there are many memory-bound operations in LLM inference, e.g., layernorm and attention mechanisms. TensorRT-LLM, DeepSpeed, and similar frameworks leverage CUDA customization to optimize these operations, resulting in speed improvements. Tri Dao et al.~\cite{dao2022flashattention} introduce Flash Attention, which performs attention calculations on SRAM with higher bandwidth to reduce memory access overhead. Building upon this, they further designed Flash AttentionV2~\cite{dao2023flashattention2}, Flash Decoding, and Flash Decoding++~\cite{hong2023flashdecoding++}.

Batching requests can effectively enhance the throughput of LLM systems. The most commonly used method is iteration-level scheduling, also known as continuous batching, introduced by ORCA~\cite{yu2022orca}. In this approach, during the autoregressive decoding process, at each step, the system dynamically assesses whether requests have been completed and whether new requests can be added, improving the overall throughput.

Memory management is also a critical bottleneck affecting the throughput of LLM systems. This is because the autoregressive encoding process in LLM generates a significant amount of KV cache, leading to the issue of memory fragmentation. Techniques such as recomputation and dynamic offloading can reduce peak memory usage, thereby increasing the system's request capacity. However, they come with additional computational or memory access overhead. PageAttention, introduced by vLLM, manages the KV cache in blocks, inspired by the paging techniques in operating systems. It effectively reduces memory fragmentation issues but still inevitably involves some memory wastage.

\section{Conclusion and Future Work}
\label{sec:conclusion}
% 0.25 page
In this paper, we focus on optimizing goodput under SLA guarantees and introduce a novel Past-Future scheduler that consists of looking to the future and reflecting on the past. It not only considers memory usage at each future time step but also leverages the historical request output length distribution to dynamically predict the output length of requests in the running batch, maximizing memory utilization and minimizing harmful request evictions. Additionally, we present LightLLM, a high-performance LLM serving framework. With the proposed Past-Future Scheduler, LightLLM can be deployed on a cluster with thousands of GPUs to support a variety of real-world LLM services.

For future work, we believe the proposed scheduler also has significant potential for the scenarios of dynamic service instance availability and fluctuating workloads. Based on its ability to accurately estimate the memory demand of each running batch, it can forward requests to underutilized services based on the current SLA constraints and service load, aiming to ensure that each service reaches full capacity. Therefore, when the request workload changes, it can achieve timely machine scaling with minimal request migration overhead.
\section*{Acknowledgements}
We thank all reviewers for their insightful comments. This work is supported by the Beijing Municipal Science and Technology Project (No. Z231100010323002), the National Natural Science Foundation of China (No. 62322201, U23B2020, 62306025, 92367204), CCF-Baidu Open Fund and the Fundamental Research Funds for the Central Universities (JK2024-58).

\bibliographystyle{ACM-Reference-Format}
\balance
\bibliography{references}

\end{document}